\documentstyle[11pt,aaspp]{article}

\newcommand{\etal}{{\it et al.\ }}   
\slugcomment{Submitted to Astrophysical Journal} 
\lefthead{} 
\righthead{}

\begin{document}

\title{A new catalog of photometric redshifts in the Hubble Deep Field
\footnote{Based on observations taken with the NASA/ESA Hubble Space  
Telescope, which is operated by AURA under NASA contract NAS5-26555 and 
at the Kitt Peak National Observatory, which is operated by AURA under 
cooperative agreement with the NSF.}}

\author{Alberto Fern\'andez-Soto\altaffilmark{2,3}, Kenneth M. Lanzetta
\altaffilmark{3}, and Amos Yahil\altaffilmark{3}}

\altaffiltext{2}{Department  of Astrophysics  and  Optics,  School  of
Physics,  University of New South  Wales, Kensington-Sydney, NSW 2052,
Australia}

\altaffiltext{3}{Department of Physics and Astronomy, State University
of New York at Stony Brook, Stony Brook, NY 11794--3800}

\date{}

\newpage

\begin{abstract}

  Using the newly available  infrared images of  the Hubble Deep Field
in the $J$,  $H$, and $K$ bands  and an optimal photometric method, we
have refined a technique to estimate the redshifts of 1067 galaxies. A
detailed comparison of our results with the spectroscopic redshifts in
those cases  where the latter are  available shows that this technique
gives very good results for bright enough objects ($AB(8140) < 26.0$).
From  a  study  of the  distribution   of  residuals ($\Delta   z_{\rm
rms}/(1+z)  \approx  0.1$ at   all   redshifts) we  conclude  that the
observed errors are mainly due to cosmic variance. This very important
result  allows for the  assessment   of errors  in  quantities to   be
directly or indirectly measured  from the catalog.  We present some of
the statistical properties of the ensemble of galaxies in the catalog,
and finish  by presenting a list of  bright  high-redshift ($z \approx
5$) candidates  extracted  from   our catalog, together   with  recent
spectroscopic redshift determinations confirming  that two of them are
at $z=5.34$ and $z=5.60$.

\end{abstract}


\newpage

\section{Introduction}

  The  Hubble Deep Field   (HDF)  image (Williams \etal  1996)  is the
deepest view of the universe yet obtained.   The image was acquired by
the Hubble Space Telescope (HST) using the Wide Field Planetary Camera
2  (WFPC2)  over a  period of  ten consecutive  days in December 1995,
yielding unprecedented sensitivity   and angular resolution over a  $3
\times 3$   arcmin$^2$ portion  of  the sky.  The image  was  acquired
through  the F300W, F450W,  F606W,  and  F814W broad-band filters  (of
central wavelengths near 2940, 4520, 5940, and 7920 \AA), allowing for
the  possibility of assigning redshifts   to  objects detected in  the
image  on  the  basis of    broad-band  photometry.  This  so   called
photometric redshift   technique was originally   applied to  galaxies
detected in other fields by various authors (see, e.g., Loh \& Spillar
1986;      Connolly  \etal     1995),   with   remarkably   successful
results.  Several groups have by now  applied the photometric redshift
technique to galaxies   detected in the  HDF,  each using a   slightly
different  approach, and  it  was  recently verified  that all  groups
obtain    essentially correct   results  in  the    limit of   bright,
moderate-redshift  ($z   < 1.4$) galaxies  (see   Hogg  \etal 1998 and
references therein).

  In   our first   work   (Lanzetta, Yahil  \& Fern\'andez-Soto  1996,
hereafter LYF), we described  our version of the  photometric redshift
technique. Using  a  maximum-likelihood analysis  applied to   precise
photometry  and  a selection of  galaxy  spectral templates (including
templates of E/S0, Sbc, Scd, and Irr galaxies), we presented a catalog
of 1683 objects  in the HDF complete  to magnitude $AB(8140) =  28.0$.
For each object, we  listed a position, magnitude, estimated redshift,
and estimated spectral type. We showed  that the redshifts were mostly
reliable, in the  sense that more than  85\% were  essentially correct
($\Delta z_{\rm   rms}  \approx   0.15$)   when  compared   with   the
then-available  redshifts   determined  by   means  of   spectroscopic
observations. Nevertheless, some issues were raised about the validity
of  the   photometric  redshift   technique,  mainly   concerning  the
possibility of  confusion    between  high-redshift   galaxies     and
heavily-reddened low-redshift galaxies.

  Since  our  first work,  infrared images  of   the HDF were acquired
through the $J$ (1.2 $\mu$m), $H$ (1.65  $\mu$m), and $K$ (2.2 $\mu$m)
broad-band  filters (Dickinson \etal 1998).  Although these images are
not as deep as the  optical images, they  are of great utility for the
photometric analysis because they  augment  by more  than a factor  of
three the available  wavelength coverage.  This additional information
is important for   two reasons: First,  it can   help to resolve   the
redshift degeneracy   observed  in  some   cases  by  LYF---where  the
interpretation of breaks   in the spectra  of  some  galaxies  is made
difficult because low-redshift  Balmer (and other) discontinuities are
confused with   high-redshift Lyman  discontinuities.  Second, it  can
help to  distinguish  high-redshift  galaxies  from   heavily-reddened
low-redshift galaxies by measuring  or constraining the observed-frame
infrared spectral energy distributions.

  The infrared images differ  qualitatively from the optical images in
that the  spatial resolution of the  infrared images is about an order
of  magnitude   lower than   the  spatial   resolution of  the optical
images.  For this reason, it is  essential that the  photometry of the
infrared images  be performed in a very   careful way, taking  as much
profit as  possible of the valuable information  that has already been
extracted  from the optical images. Here  we use the optical images to
create  model spatial profiles that  are fitted to the infrared images
in order to measure optimal infrared fluxes and uncertainties. Even in
cases where the galaxies  are not bright enough  to be detected in the
infrared images,  we are able  to establish  firm upper limits  to the
infrared  emission.   These limits  have  direct  consequences on  the
redshift likelihood  functions,   and  they also  strictly   limit the
reddening that can be affecting the galaxies under consideration.

  In this paper, we  present our new analysis.  We explain the changes
we  have made to  our technique  in  order to incorporate the infrared
images, and we   present a new catalog  of  photometric redshifts. The
catalog  includes  optimal   infrared fluxes---or  upper  limits---and
uncertainties  as well   as  photometric redshift  estimates  for 1067
objects.  We also  present a    comparison of  our  results with   new
spectroscopic observations that  have become available since our first
work.

  The organization of this paper is as follows: Section 2 presents the
data.  Section 3    describes the  object   detection and   photometry
procedures,    with particular     attention  to   the  infrared  flux
measurements. The method used   to obtain the estimated redshifts   is
presented  in  \S\ 4.The  final  catalog is   presented  in \S\  5 and
compared   to  spectroscopically  determined redshifts  in   \S\ 6. We
discuss some  relevant properties of the  distribution  of galaxies in
\S\ 7, and review our main conclusions in \S\ 8.

\section{Data}

  The optical images were acquired by HST in December 1995 using WFPC2
and the F300W, F450W, F606W, and  F814W filters (Williams \etal 1996).
Reduced versions of  the images were  made available for public access
in January 1996. The  physical format of  the optical images is  $4096
\times  4096$ pixel$^2$ at a plate  scale of 0.04 arcsec pixel$^{-1}$,
which   corresponds  to an   angular area   of   $\approx 3  \times 3$
arcmin$^2$.   Throughout this  work,  we  refer  to  coordinates  with
respect  to the Mosaic version  of the images,   which were created by
Mark Dickinson and Richard Hook from the  Version 2 images provided by
the  STScI team, and  to the filters F300W, F450W,  F606W and F814W as
$U$, $B$, $V$, and $I$ respectively.

  The infrared images were obtained by  Kitt Peak National Observatory
4 m  telescope in April 1996  using the IRIM  camera and standard $J$,
$H$,  and $K$ filters (Dickinson \etal  1998). Reduced versions of the
images were  made available for  public access  in November 1996.  The
physical format of the infrared images is $1024 \times 1024$ pixel$^2$
at a plate scale of 0.16 arcsec  pixel$^{-1}$, which corresponds to an
angular area  of $\approx 3 \times 3$  arcmin$^2$. The infrared images
are registered to the same coordinate grid as the optical images.

\section{Object detection and photometry}

\subsection{Object detection}

  We use the SExtract source extraction  code (Bertin \& Arnouts 1996)
to detect objects  in the F814W  image. Details  of the procedure  are
described by  LYF96.  At this stage,  the  object  list contains  2323
objects with {\it best}  total $AB(8140)$ magnitudes down to  $\approx
30$. Total magnitudes  labeled by SExtract as  {\it best\/}  are first
measured  as aperture magnitudes and  then corrected for the effect of
neighboring   objects and  the    flux  lost in   pixels  outside  the
aperture. We   will use  these  {\it best}   magnitudes to define  the
completeness limits  of  our catalog as well  as  in the  studies that
depend on magnitudes of individual objects (like the galaxy counts and
other  studies sketched   in   Section  \ref{discussion}),   but   the
calculations described in this and the following Section will be based
on aperture magnitudes as define in Section \ref{photom}.

  It is known that the edges  of the Wide  Field Camera images as well
as the Planetary Camera image  are of poorer  quality than the bulk of
the Wide Field Camera  images. In  our  previous work, we  trimmed the
image  borders to avoid contamination   by  spurious sources. In   the
present work,  we instead define two  different zones of two different
depths in order not to exclude some bright objects that are located in
the  previously   trimmed  area,  some   of  which  have spectroscopic
redshifts available.

  The inner part of the image (hereafter zone 1,  or Z1) is the deeper
zone, which is cut to a magnitude limit of $AB(8140)  = 28.0$. To this
limit, there are 946  galaxies  spread over an   angular area of  3.92
arcmin$^2$.  The  outer   part  of the  image   (hereafter  zone 2, or
Z2)---including the  external borders of the  Wide Field Camera images
and the Planetary Camera image---is  the shallower zone, which is  cut
to a magnitude  limit of $AB(8140)  = 26.0$. To  this limit, there are
121 galaxies spread over an angular area  of 1.39 arcmin$^2$. Zones Z1
and Z2 are defined in Figure \ref{zones}.

\subsection{Photometry}
\label{photom}

  To perform the optical photometry, we  repeat the analysis described
by  LYF. This analysis  is based on the  definition of an aperture for
every object by means of the segmentation map given by the application
of SExtract to the F814W image.  The segmentation map marks the pixels
of every  object detected in the  F814W image, which corresponds to an
isophotal aperture of surface brightness limit $\mu_{AB(8140)} = 26.1$
arcsec$^{-2}$.

  Given the apertures, we calculate the observed  flux of every object
in every optical image by summing  the fluxes within the apertures and
subtracting the   backgrounds  determined in  boxes   surrounding  the
objects.   To assess errors in the   flux values, we  calculate the $3
\times  3$  data  covariance matrix  of  every image,  because,  as is
described by Williams \etal (1996), the drizzling algorithm applied to
the images  to  improve the  spatial resolution produces  correlations
between adjacent pixels (stronger between immediate neighbors and less
important, but still significant, in diagonal directions).  We account
for this effect in the calculation of the photometric errors using the
data covariance matrix, that has the net effect of increasing the flux
errors  that would be measured  otherwise. This increase is determined
essentially  by  the values  of   the correlation between  neighboring
pixels, with  a correction that  depends on the shape  and size of the
object---number of pixels within the measured aperture.

  Using apertures fixed   by the F814W  image  assures that the   same
portion of every object is measured in every filter, hence this method
is preferable to any other variable-aperture  technique for this work.
But this  approach  cannot  be easily adapted   to photometry  of  the
infrared images.  The difference  between the infrared images  and the
optical images  is, of course,  that the ground-based  infrared images
are of significantly   lower spatial resolution than   the space-based
optical images.  To account for this difference,  we apply a procedure
that yields essentially   optimal infrared photometry  of every object
detected in the optical images.

  We make the  single assumption that  the spatial profiles of objects
detected in the F814W image---which  is observed at the extremely high
spatial resolution  provided by HST---are  good tracers of the spatial
profiles at observed-frame infrared  wavelengths.  This  assumption is
plausible  for the following reasons:    For objects of low  redshift,
observed-frame infrared wavelengths  correspond to rest-frame infrared
or red  wavelengths, at which  galaxy morphology   is not a  sensitive
function of wavelength.  For objects of  high redshift, typical galaxy
angular sizes  are much less than  the seeing profile  of the infrared
images  (LYF),  so the objects   are almost always  unresolved  by the
infrared images.  Together,   these effects  make the  above-mentioned
assumption reasonable  in  all  cases  of   interest for  the  present
analysis. Moreover, $BVRIHK$  observations  of spiral galaxies   by de
Jong   \& van der     Kruit (1994)   and   de  Jong  (1996)   show  no
significatively different profiles in the optical and infrared bands.

  First we determine  the shapes of  the point spread functions of the
infrared images by fitting Gaussian  profiles to stars detected in the
images.  The variation of the  point spread functions with position on
the field  is taken into account by  fitting  quadratic polynomials to
the point  spread  function FWHM as  functions of  the distance to the
center  of   the   field.   Table   \ref{psf}  shows  the   polynomial
coefficients obtained    from  these  fits, from    which   it can  be
appreciated  that this effect must be  taken  into account in order to
perform accurate photometry.

  Next, we   convolve the  F814W   image of  every  object   with  the
appropriate point spread functions in  order  to obtain models of  the
infrared  spatial  profiles in   the  three  infrared bands. We   call
$P_i^{(b)}(m,n)$  the model profile of  object $i$ at pixel $(m,n)$ in
infrared  band $b$, normalized    to  unit  total flux.  Using     the
$P_i^{(b)}(m,n)$, we create model  images $M^{(b)}$, which  are linear
sums  of the profiles  of every  object in band  $b$.  The  profile of
every object is multiplied by a flux $F_i^{(b)}$ in band $b$, which is
a parameter  to be    fitted.  The  individual profiles  are    summed
according to
\begin{equation}
M^{(b)} (m,n) = \sum_{i=1}^{N_{\rm obj}} F_i^{(b)} P_i^{(b)}(m,n).
\end{equation}

  In  order to achieve  maximum accuracy  in the  determination of the
fluxes,  it is also necessary  to obtain a  very good  estimate of the
background and  the noise   in every  image  and of   its  statistical
characteristics.   To obtain this  estimate  for the $i$th object,  we
calculate   the local background $B_i^{(b)}$   and  $3 \times 3$  data
covariance matrix $\bar \sigma_i^{(b)}$    using all pixels in a   box
surrounding the object  that are  not identified  as belonging to  any
source---including  of course   the  object  under  study.   The  data
covariance  matrix  is normalized so  that  its central  element $\bar
\sigma_i^{(b)}(2,2)$ is equal   to $\sigma_{\rm loc}^2$, the  standard
deviation of the local distribution of noise values around the source.
We then define an {\it effective} $\sigma$ for that area to be
\begin{equation}
\sigma_{{\rm eff}_i}^{(b)} = \left[ \sum_{i_1,i_2=1}^{i_1,i_2=3} \bar
\sigma_i^{(b)}(i_1,i_2) \right]^{1/2}, 
\end{equation}    
which is actually a slightly conservative overestimate of the real 
error.

  The model image  $M^{(b)}$ is then  fitted to the observed  infrared
image $I^{(b)}$ by means  of a $\chi^2$ linear minimization technique,
rendering the equation
\begin{equation}
\chi^2=  \sum_{m,n}\left[\frac{I^{(b)}(m,n)-B^{(b)}(m,n)-M^{(b)}(m,n)}
{\sigma_{{\rm eff}}^{(b)}(m,n)}\right]^2.
\end{equation}

  This requires the resolution of a $N_{\rm obj}^2$ matrix which is 
very sparse, because the only non-zero elements arise at pairs of indices
whose related objects overlap. The set of $N_{\rm obj}$ equations that 
minimizes $\chi^2$ is
\begin{equation}
\frac{\partial \chi^2}{\partial F_i^{(b)}}=0 \;\;\;(i=1,\ldots,N_{\rm obj}),
\end{equation}
which leads to the Hessian matrix of the system:
\begin{equation}
A_{ij}^{(b)} = \frac{\sum_{m,n} P_i^{(b)}(m,n) P_j^{(b)}(m,n) }
{\sigma_{{\rm eff}_i}^{(b)} \sigma_{{\rm eff}_j}^{(b)}},
\end{equation}
and the right hand side of the equation:
\begin{equation}
R_i^{(b)} = \frac{\sum_{m,n} P_i^{(b)}(m,n) [I^{(b)}(m,n)-B_i^{(b)}]}
{(\sigma_{{\rm eff}_i}^{(b)})^2}.
\end{equation}
In both definitions the  indices $(m,n)$ span  the entire  image.  The
result of this  procedure is the set of  fitted fluxes $F_i^{(b)}$ and
associated  uncertainties   $\sigma_{{\rm  F}_i}^{(b)}$  for    $i   =
1,...,N_{\rm obj})$.

\subsection {Validation tests for the IR photometry}

  As the procedure we have sketched in  this Section is rather complex
and, as far as we know, has not been  presented before, we consider it
necessary to perform some tests on the results  to assess the validity
of the technique.

  A first idea on the goodness of the achieved fit  can be obtained by
simply having a look at Figure \ref{image}, which shows a piece of the
$K$  band image, together with the  corresponding F814W image, the $K$
band model  as fitted  by  our procedure   and the residual   image. A
detailed analysis of the residual image suggests that the infrared PSF
might not  be  perfectly Gaussian,  which  may produce   errors in our
photometry.  We   have checked that   these  errors  are  smaller than
$\Delta[AB(22000)] \approx  0.1$ for the brightest objects ($AB(22000)
\approx 18$) and smaller  than $\Delta[AB(22000)] \approx 0.2$ for the
faintest ones  (those with  $AB(22000) \approx  24$). In  general this
systematic error is smaller than the photometric error for all objects
with  $AB(22000) \gtrsim  23$---which  is  approximately 90\%  of  the
catalog.

  We now  show two tests  that are  very different  in nature and will
help us to establish the accuracy of the  measured IR fluxes and their
associated noise estimates.

\subsubsection {Infrared photometry of faint objects}

  Because the optical images  are of much  higher quality (in terms of
depth and spatial resolution) than the infrared  images, we can expect
a  very significative fraction of the  objects to have infrared fluxes
measured  by  our technique that  are    compatible with zero.  Having
measured a flux $F_i$ and an associated error $\sigma_i$ for every one
of the objects in any infrared image, we can form the quotients
\begin{equation}
 q_i = F_i / \sigma_i
\end{equation}
which tell us how significantly different from zero every flux is that
we measure on the  image.  Excluding the objects  with high  values of
$q_i$, we can concentrate  only  on the above-mentioned objects  whose
flux is compatible with zero.

  We can  use this sample of objects  to check that our error analysis
has been correct. If  that is the case, the  distribution of values of
$q_i$ must have two components: First, a normal distribution with mean
value $\bar q = 0$ and variance $\sigma_q  = 1$, that accounts for the
fact that a  fraction    of  the  flux measurements  are    completely
indistinguishable  from zero. Second, a group  of positive values that
makes the $q>0$ side broader, corresponding  to measurements of fluxes
from  weak sources. The  relative  importance of these two  components
will depend on  the relative depth  and spatial resolution of  the two
images involved (in our  case, F814W for the  detection of the objects
and the infrared image for the flux measurements). Other properties of
the images (like the flat-fielding and putative background variations)
are not a limiting factor in this analysis, as the quality of both the
HST and Kitt Peak  images is very  good. We  decided, in any  case, to
estimate the background and covariance  matrix locally for each object
in each band, which makes our analysis even more robust in that sense.

  In order to check whether  the two components described above appear
in our  flux distributions and to  measure their relative importances,
we  show in Figure  \ref{errors} the  distributions of  $q_i$ for  all
three infrared bands, and superimposed on them a Gaussian distribution
($\bar q  = 0$,  $\sigma_q =  1$)  whose  only free parameter   is the
normalization  level.   This normalization   is calculated to  fit the
number of  objects {\em in the negative  part of the distribution}, to
avoid the influence of those   weak positive flux measurements on  the
positive side of the distribution.

  Looking at the $q<0$ distributions we can see  that there is perfect
agreement between  the expected and   observed distributions. This  is
proof  that the infrared photometric errors  are Gaussian and that our
estimation  of their values is  likely correct. The  other side of the
distribution  ($q>0$) is    broadened---as expected---because of   the
accumulation of positive  measurements that  correspond to very  faint
sources.

\subsubsection {Distribution of the residuals}

  In order to check that the model light profiles that we are using in
our fit  are correct, we can  study  the distribution of  the residual
intensities in the image that  results from subtracting the model from
the original image. First,  each of the infrared  images is divided in
two different sectors.   The first sector   contains all those  pixels
whose values in the model image are  significantly different than zero
(using a $1\sigma_{\rm loc}$  threshold, with $\sigma_{\rm loc}$ being
the standard deviation  of   the  local distribution of     background
values).  The second sector contains the rest of the points, i.e., the
ones   that---according   to  the   model---contain essentially   only
background noise.

  Figure  \ref{residuals} shows in the upper  panels the histograms of
the pixel  intensity distributions---normalized to unit area---in both
sectors for   all  three  bands.  They   are obviously  different,  as
expected, with  the intensities in  the first sector (continuous line)
being higher  than the ones in the  second  (dotted line). This merely
shows the fact that the pixels corresponding to the sources, which are
those  with high intensities, are all  in the  first group. The dashed
lines (corresponding to  background pixel intensity distributions) are
all consistent with Gaussian noise.

  We can now  look at the  histograms of the intensities corresponding
to the same pixels in the residual  images (i.e., $I^{(b)} - M^{(b)}$,
lower panels of the same Figure). In this case, both distributions are
seen  to have the  same mean  value. This  indicates that the residual
source intensities  average to  zero   which means that   our measured
fluxes are not biased  either towards low  or high values. However, we
observe in the same panels that the variance  around zero is different
in the intensity distributions for the  first and second sectors. This
is expected, as it   is due to the  fact  that we are subtracting  the
intensities corresponding to the models, but of course we cannot avoid
the fact  that the  noise (essentially Poissonian  on the  images)  is
higher where the signal used to be higher.

\section{Redshift estimation}

  We estimate the redshift of every object detected in the F814W image
from the photometric data using the technique described by LYF, except
that here we use seven optical and  infrared photometric bands whereas
LYF used only four optical photometric bands.   The idea is to compare
the observed spectral energy distributions  (SED) with a grid of model
galaxy SEDs   (including  the effects  of  intrinsic  and  intervening
absorption) ranging in spectral type and redshift.

\subsection{Selection of the model spectra}

  We use as  basic galaxy spectra the  four spectra templates given by
Coleman,   Wu \& Weedman  (1980).    They include four  different {\em
observed}  galaxy types  (E/S0, Sbc,   Scd, and   Irr) and cover   the
wavelength range 1400-10000  \AA.  We extrapolate these  templates  at
ultraviolet  wavelengths  to 912 \AA\  using  results  of Kinney \etal
(1993) and at infrared  wavelengths  up to  25000  \AA\ using spectral
evolutionary models of Bruzual   \& Charlot (1993).  We show  in Table
\ref{galparam} the parameters we  have used to construct the  infrared
extensions of the  templates,  as suggested  by Pozzetti, Bruzual   \&
Zamorani (1996). Figure \ref{sed} shows the four spectra used as input
models. We   do   not   include  Lyman   $\alpha$   emission   in  our
templates. This is due to the large  spread of Lyman $\alpha$ emission
line   equivalent widths   observed  in   moderate  to   high-redshift
galaxies---ranging    from  negative  (heavily  absorbed  profiles) to
$\approx$  60 \AA. Even the  largest   quoted equivalent widths  would
produce only minor effects in our broad-band photometry except for the
putative  case of very  strong Lyman $\alpha$  lines in galaxies at $z
\approx 4.1$ or  $z \approx 6.0$, when  the lines would occupy the red
end of the    F606W or F814W  filters  respectively.  These cases  may
deserve future studies,  specially as the number  of spectroscopically
confirmed $z > 4$ galaxies is increased.

  Intrinsic  neutral hydrogen  absorption   is incorporated  into  the
models  by assuming that   galaxies  are optically  thick  to ionizing
radiation below the  Lyman Limit (912 \AA\ in  the galaxy rest frame).
This assumption has been verified by Leitherer \etal (1995) for nearby
starburst galaxies  and   has  been  checked  in those  objects  where
spectroscopic redshifts  are available.  Intervening  neutral hydrogen
absorption is incorporated into the models (as a function of redshift)
by  adopting the average  Lyman  $\alpha$ and Lyman $\beta$  decrement
parameters $D_A$  and $D_B$, as  measured  by  Madau (1995)  and  Webb
(1996).  Given that these measurements  extend only over the  redshift
range $0<z<6$,  we extrapolate the values to  higher redshifts using a
simple polynomial  fit passing through Webb  (1996) points $D_A(z=6) =
0.951$, $D_B(z=6) = 0.961$ and forcing $D_A =D_B =1$ for $z = 8$.

  We   do not include evolutionary  corrections   in the model spectra
because such  corrections are very uncertain  and would add additional
parameters to the model---including, for example, the galaxy ages.  We
believe, moreover, that the adopted galaxy  spectra span a wide enough
range of spectral   properties to account  for  evolution in  a purely
passive way.  However, it is certainly true that distant galaxies with
spectral   properties  significantly different   from  all present-day
galaxies would  render a problem for  our method.  Such cases have not
been seen in  observations  of high-redshift  galaxies  so far  (Cohen
\etal 1996;  Lowenthal \etal  1997;  Zepf, Moustakas, \& Davies  1997;
Steidel \etal  1996).  In  the particular  case of very  high-redshift
galaxies, their main spectral feature is  the utter lack of flux below
$912  (1+z)$ \AA\ and almost complete  lack of flux below $1215 (1+z)$
\AA.  In these cases, other  features of the  spectra turn out to have
close to   negligible   effect  on the   redshift estimates,   as   is
demonstrated by the  fact that  {\em at  high redshifts}  our  fitting
procedure gives almost the same goodness  of fit and redshift estimate
for any of the three bluer models (the  elliptical galaxy is ruled out
because   of the  large excess of   flux  that should  be observed  at
infrared wavelengths).

  The redshifted  and absorbed spectrum  of every galaxy model is then
integrated with  the  throughputs of   the HST and   Kitt Peak filters
(including instrumental  throughputs, made  available through  the HST
and Kitt  Peak  databases) in order to   obtain the  expected measured
fluxes for every galaxy type and redshift.

\subsection{Redshift likelihood functions}

  Once the expected  fluxes in all seven bands  are calculated for all
the  models, we construct the  redshift likelihood functions according
to  the following scheme.   Assuming  that the flux  uncertainties are
normally  distributed,  the likelihood $L(z,T)$ of  obtaining measured
fluxes $f_i$  with   uncertainties  $\sigma_i$   given modeled  fluxes
$F_i(z,T)$ for a  given spectral type T at  redshift $z$, with  a flux
normalization $A$ over the seven filters ($i=1-7$) is
\begin{equation}
L(z,T)=\prod_{i=1}^{7} \exp\left\{-\frac{1}{2}\left[
\frac{f_i-AF_i(z,T)}{\sigma_i} \right] ^2 \right\}.
\end{equation}
This likelihood function is then  maximized  with respect to  spectral
type and normalization $A$ to  obtain the redshift likelihood function
$L(z)$.  The  redshift likelihood   function is then   maximized  with
respect to redshift to yield  the  maximum-likelihood estimate of  the
redshift, $z_{\rm phot}$.

  The likelihood functions $L(z)$   should not only give estimates  of
$z$,   but   also some confidence   level  around  the best-fit value.
Unfortunately, the  problem is not a  well-defined one  because of the
cosmic  variance involved  in the  selection  of a finite---and in our
case small---number of spectral templates.  Because of this effect, we
cannot  in general define  real confidence limits, but  only use in an
orientative  way the likelihood   functions to distinguish those cases
where the  value  of $z_{\rm phot}$  is  susceptible  of error---e.g.,
those ones  where the  function $L(z)$ is  multi-modal.  For the  same
reason, individual values  of  $L(z)$  corresponding to  the   maximum
$z=z_{\rm phot}$  are not  good  measurements of the  goodness of fit.
For example,  they are completely  dominated by the cosmic variance in
those  cases   where   the sources  are  bright    enough to  make the
photometric uncertainties very small, which renders very low values of
$L(z)$ even for apparently  excellent  fits. Nevertheless, as we  will
discuss  in the next  Section, photometric errors dominate over cosmic
variance  at  magnitudes $AB(8140)   \gtrsim 26$. In  this regime  the
likelihood functions do   handle  the errors properly,  so  confidence
limits on $z$ can be obtained from them in the usual way.

\section{The catalog}

  We present in this  Section the last steps  performed on the  object
list in order to reach the final catalog. These include the correction
of those single   objects  that were broken  into  separate  pieces by
SExtract and the elimination of stellar objects in the HDF.

\subsection{Correction of deblended objects}

  Automatic deblending of object images has been known for a long time
to be a very  difficult task. Although SExtract does  a very good work
in  creating   the initial  list from  the   F814W  image,  a  careful
examination of  the segmentation maps  shows that some groups  of very
close     images,  that  might well  belong    to   the  same physical
object---however difficult this might  be to define---have been broken
into separate pieces in our catalog. We decide  to attack this problem
{\em  after} the photometric analysis has  been  performed, because in
this way we   can use  the  color  information  ---together with   the
redshift estimate---of each  of the pieces in   every group to  decide
whether or not to add them together into one single object.

  We select from  the catalog all  those  pairs (or larger groups)  of
objects whose centers lie on the image at distances smaller than 2 arc
seconds---or 3 arc seconds if any of the objects has $AB(814) < 24.0$.
These  groups are analyzed  in a one-by-one basis,  and in those cases
where the colors of the sub-clumps are  found to be compatible and the
images suggest that there is only one source, we add the corresponding
fluxes together  in an optimal  way and repeat  the calculation of the
redshift likelihood for the resulting single object.

  The total number of  groups analyzed in  this way was over 200, most
of them  pairs   and also some  triads   and even larger   groups.  In
approximately half  of the cases  the sub-objects  were found to share
similar colors, and hence were added together as a single entry in the
catalog.  This   procedure makes    our   catalog less  prone  to  the
over-counting  of faint small objects  (see Colley \etal 1996) and the
numbers involved show that  it can indeed pose a  problem at  the high
resolution achieved with HST imaging.

\subsection{Stars in the HDF images}

  We have decided not to include a stellar spectrum in our set of SEDs
in this work. We  take this  decision based on  two facts.  First, our
previous  analysis (LYF) did include  a M star   template and we found
that  none of the objects  was better fitted by it  than by any of the
galaxy  templates. Second,  and  most  important, several compilations
have been published of point-like sources in  the HDF (Flynn, Gould \&
Bahcall 1996, Elson,  Santiago \& Gilmore  1996,  M\'endez \etal 1996,
Cowie 1997). Using them we have chosen  all the objects that appear as
stellar at  least  in  one of the   references---20 objects  in total,
including an obvious star not included in them possibly because of its
position  being   too  close to  the  border  of   the  images---, and
subsequently eliminated them from our catalog. Table \ref{stars} lists
the sources that have been eliminated.

\subsection{The final catalog}

  The complete catalog contains 1067  entries.  The inner area of  the
HDF images (Z1) contains 946 galaxies over an area of 3.92 arcmin$^2$,
reaching down to  magnitude $AB(8140) = 28.0$.  The outer area (Z2) is
less deep, containing 121 objects  with magnitudes down to $AB(8140) =
26.0$ over an area of 1.39 arcmin$^2$.

  Table 4 lists the coordinates,  {\it best\/} $AB(8140)$ magnitudes,
redshifts  (both spectroscopic,  if  available,  and photometric)  and
best-fit spectral types for every galaxy in the catalog, together with
the  corresponding ID in the  Williams \etal (1996)  catalog. Table 5
lists the {\it aperture} fluxes of every object in all seven $UBVIJHK$
bands together  with their  associate  1$\sigma$  errors in   nJy. The
relations between $AB$ magnitudes and flux are given by:
\begin{equation}
AB_{\lambda}=-2.5 \log f_{\nu}({\rm  erg\ } {\rm cm}^{-2} {\rm s}^{-1}
{\rm Hz}^{-1}) -48.60 = -2.5 \log f_{\nu}({\rm nJy})+31.40.
\end{equation}

  Tables 4 and 5 (both in HTML format)  are available in the {\it on
line} edition,   and include  links from  every  object  entry  to its
individual  data   page. Both  Tables, together  with   the catalog in
electronic form  and  a  clickable    map  to help with   the   object
identification, are available upon  request from the authors, and also
at the   following   WWW address:  {\tt  http://bat.phys.unsw.edu.au/}
$\tilde{ }$ {\tt fsoto/hdfcat.html}.

\section{Comparison with spectroscopic redshifts}

  As was mentioned before, there are $\approx 100$ galaxies in the HDF
that have had their redshifts  already determined through spectroscopy
(Cohen \etal  1996, Cowie 1997, Steidel \etal   1996, Zepf \etal 1997,
Lowenthal  \etal 1997). As is  explained in Lanzetta, Fern\'andez-Soto
and    Yahil   (1997),   there   have  been    problems  regarding the
identification  of    sources in  the  HDF  that  render  some  of the
spectroscopic redshifts  uncertain.    After a  very  careful    check
procedure, we have obtained a list  of reliable redshifts and enriched
it  with some data not yet  published (kindly provided  by Spinrad and
Steidel and Dickinson, see also Dickinson \etal 1997).

  This list contains   108   redshifts. The comparison    between  the
photometric and spectroscopic values is  shown in Figure \ref{spvsph}.
These are the main features that can be extracted from it:

  1. There are  two   obvious  errors corresponding to   high-redshift
  1. galaxies
($z>2$)  to which we have assigned  low-redshift values. Their numbers
in the catalog are 687  ($z_{\rm  sp}=2.931, z_{\rm phot}=0.280$)  and
1044  ($z_{\rm sp}=2.008,  z_{\rm   phot}=0.040$). In both cases   the
redshift  likelihood functions show secondary  peaks close to the true
value (3.120  and 2.280 respectively),  but the SEDs are better fitted
by the wrong redshift. We have made some  attempts to change the model
SEDs in order to avoid these problems, but any change in the templates
that leads  to the correction of this  problem causes other fits to go
wrong.

  2. The rest of the high-redshift sample (27 objects, 93\% of total
galaxies with $z>2$) shows  an acceptable distribution in the  $z_{\rm
sp} -z_{\rm  phot}$ plane, with  an  {\it rms} dispersion $\Delta_{\rm
rms}z \approx 0.45$. Over 70\% of the total  of objects (21 out of 29)
have  a deviation from  the correct  value which  is less  than 0.5 in
redshift units.  A general trend is  noticed, however, to assign lower
redshift values to the galaxies  in the redshift range $z=2.0-2.5$. We
have tried but failed  to find an explanation and  a solution to  this
problem. It is not  related to the  absence of Lyman $\alpha$ emission
in our  spectral templates, as  there is  no correlation between Lyman
$\alpha$  equivalent   width   and the  measured   value  of  $(z_{\rm
sp}-z_{\rm phot})$.  Perhaps the inclusion  of evolution in the models
could  palliate this problem somehow, at  the  price of including more
parameters in the fit.

  3. At low redshift ($z<1.50$) the situation is essentially perfect,
with a dispersion $\Delta_{\rm rms}z = 0.13$. If  the 5 points with $|
z_{\rm   sp} - z_{\rm  phot}  | > 0.20$---which   represent 6\% of the
sample---are   eliminated, the  dispersion   goes down to $\Delta_{\rm
rms}z = 0.09$.

  It is  worth remarking that the  scatter of the high redshift values
is comparable   to the one  of  the low  redshift ones,   if we use as
estimator  the value of $\Delta_{\rm   rms}z/(1+z)$, with $\Delta z  =
z_{\rm  sp}    - z_{\rm  phot}$.   This choice    is  justified by the
functional dependence of   most physical quantities  on redshift.  The
values of this estimator---using average redshifts 0.75 (low-redshift)
and  2.75 (high-redshift)---are then  0.07 and  0.12, respectively. In
Figure \ref{delta} we  show the average  values of this estimator when
the  data are binned  in redshift (groups of   nine or ten objects per
bin). The points mark the median value of $z$ and $\Delta z/(1+z)$ for
every bin, with the horizontal error bars showing the width of the bin
and the vertical ones  marking the 25\% and  75\% percentiles in every
bin. In order to create  this plot we have ignored  the two very large
high-redshift errors on which we already commented.

\subsection{Photometric Errors versus Cosmic Variance}

  Once the identification  of objects is  secure and the spectroscopic
redshifts  have been double-checked  to  avoid possible mistakes  (see
Lanzetta \etal 1997), two   different effects  may contribute  to  the
residuals  between  the photometric  and the  spectroscopic redshifts.
These are cosmic variance and photometric uncertainties.

  Cosmic variance influences the results because we are using a finite
set of spectral templates to fit an  otherwise continuum spectral type
space. The fact  that  our templates do  not exactly  fit all possible
spectral energy distributions  will  produce  an uncertainty in    the
fitted  redshifts. This uncertainty might be  very large if any galaxy
spectrum came up to be extremely different  from all of our models. As
mentioned  above,  this  has not    been  observed in the  spectra  of
high-redshift galaxies reported to date.

  On the  other side,  even for  those galaxies  whose spectral energy
distribution would be  perfectly fitted by one  of our templates,  the
photometric  uncertainties will  produce a residual   between real and
estimated redshift.  This   residual  is  obviously  expected   to  be
increasingly important as a function of apparent magnitude.

  In  order to evaluate the relative  importance of these two effects,
we perform the test  described here. First,  we assume our photometric
redshifts to be  exact, and the best-fit  models to be perfect fits to
the unknown underlying galaxy spectra.  In this way  we get rid of the
effect of cosmic variance, as all  galaxies in our sample become exact
copies  of  one of  our four templates.   Second,  we add random noise
(according to  the actual  flux uncertainties)  to the best-fit  model
fluxes of all objects in all bands,  and use these  fluxes as input to
our redshift-estimation  program.   All these steps  are  repeated 100
times,  aiming  to  determine  the  distribution of residuals  between
the---supposedly--- exact   redshifts and  the  simulated ones.  These
residuals are now the {\em only} product of photometric uncertainties,
as  we  have eliminated  any  cosmic variance    by assuming that  our
best-fit model represent a perfect image of reality.
 
  Figure \ref{boots} shows the distribution of residuals as a function
of apparent magnitude ({\it best} total magnitudes are used throughout
this Section) and redshift, together with the median residual for each
bin. Several conclusions can be drawn from it:

  1. Photometric error produces almost no effect on the photometric
redshifts at magnitudes $AB(8140) < 25$, independently of redshift. At
these magnitudes,  the median absolute  residual  is no  greater  than
0.04, and the distributions of  residuals are concentrated into single
peaks centered at zero residual.

  2. Photometric error produces a mild effect on the photometric
redshifts at magnitudes $AB(8140) = 25 - 26$, somewhat more pronounced
at redshifts  $z > 2$  than at redshifts  $z < 2$. The median absolute
residual  (at all  redshifts) is  only 0.04 (comparable  to the median
absolute residual  at  brighter magnitudes), but the  distributions of
residuals at  high-redshifts  are  spread into small  secondary  peaks
centered at large negative residuals  in addition to prominent primary
peaks centered at zero   residual. These secondary peaks   result from
high-redshift galaxies that are incorrectly assigned low redshifts and
contain 2.5\% of the total at redshifts $z  = 2 - 3$  and 6.1\% of the
total at redshifts $z = 3 - 4$.

  3. At magnitudes $AB(8140) > 26$ photometric uncertainties produce
an  increasingly significant effect.  At redshifts  $z  = 0 - 1$,  the
effect of photometric error at faint magnitudes is twofold: the median
absolute residuals  increase  and  long   tails  stretching to   large
positive residuals appear on the  distributions. At redshifts $z > 1$,
the effect of   photometric error at  faint magnitudes  increases  the
median absolute    residuals and produces    prominent secondary peaks
centered at large negative residuals on the distributions---the latter
again  due to the  wrong assignment of  low redshifts to high-redshift
galaxies.  In the magnitude range $AB(8140) = 26 - 27$ these secondary
peaks contain 9.7\% of the total of objects  at redshifts $z  = 2 - 3$
and 19.6\% at redshifts $z = 3 - 4$.

  4. At faint magnitudes, it is more likely that high-redshift
galaxies  will be  incorrectly    assigned  low redshifts  than   that
low-redshift galaxies will be incorrectly assigned high redshifts. For
example, at magnitudes $AB(8140) = 27 -  28$, the long tail stretching
to large positive residuals  at redshifts $z =  0 - 1$ contains 27.3\%
of  the total while the secondary  peak at large negative residuals at
redshifts $z = 3 - 4$ contains 34.8\% of the total.  This is also seen
in the values  of  the median absolute residuals,  that  are  0.26 and
0.64, respectively.

  We can  combine these conclusions  with the  results of the previous
subsection in  order  to  assess the  relative  importance  of  cosmic
variance   and   photometric uncertainties   in the  determination  of
photometric redshifts. The  low-redshift sample that has been observed
spectroscopically  contains   objects   of  magnitudes  ranging   from
$AB(8140) = 18.16$ to $AB(8140) = 24.88$, with a median of $AB(8140) =
22.55$.  We  have  just shown  that  at these  magnitudes  photometric
uncertainties  produce no  effect    on the  redshift  estimates.  The
high-redshift galaxies with spectroscopic redshifts have magnitudes in
the range  $AB(8140)  = 23.24$ to 26.75,  with  a median  of $AB(8140)
=24.75$. In this range the effect of photometric uncertainties is---at
most---mild.  We conclude   from this  analysis   that  {\em residuals
between the photometric and  spectroscopic redshifts are  dominated by
cosmic variance rather than by photometric error}.

  This  important result establishes  the  magnitude of the effect  of
cosmic variance  on the  photometric  redshifts. Specifically,  we can
attribute    the  entire    residual   between   the  photometric  and
spectroscopic redshifts to the  effect of cosmic variance and consider
the redshift range  $z  \approx 0 -  4$  spanned by the  comparison of
photometric and spectroscopic    redshifts and conclude  that   cosmic
variance  produces    a  median  absolute    residual   of   0.08,   a
$3\sigma$-clipped  {\it   rms}  residual of   0.18,   and a discordant
fraction of   $0.057$. If  the  effect of    cosmic variance does  not
increase with  decreasing galaxy luminosity---and  we can think  of no
reason   that  it  should---then  these  values   must   apply  at all
magnitudes,   not just   the   relatively  bright  magnitudes  of  the
spectroscopic limit.

 This fact    has two important   implications:  First,  at magnitudes
brighter than $AB(8140) \approx 26 - 27$, the accuracy and reliability
of the photometric redshifts is  limited by cosmic variance whereas at
$AB(8140) \ge 26 -27$ it is  limited by photometric error. Second, the
uncertainty associated  with any statistical    moment of the   galaxy
distribution  (including  the effects  of sampling  error, photometric
error, and cosmic variance) can be realistically estimated by means of
a  bootstrap resampling  technique, where  the   effect of photometric
error is simulated by adding random noise to the fluxes and the effect
of cosmic variance is mimicked by  adding (rather modest) random noise
to   the  estimated redshifts.  We expect  that   this  technique will
ultimately play an important role  in exploiting the full potential of
the broad-band photometric redshift techniques.

\section{Discussion}
\label{discussion}

  We  want  to discuss in  this  Section some of the  most interesting
properties of the  distribution of galaxies in the  HDF as revealed by
our  catalog.  A more  in-depth  analysis   is left  for  other papers
(Lanzetta, Yahil \& Fern\'andez-Soto 1998, on the feasible presence of
galaxies at $z>6$ detected  only in the  infrared images; Driver \etal
1998, on the combination   of  our catalog with galaxy   morphological
information; Lanzetta,  Fern\'andez-Soto    \&  Yahil 1998,   on   the
highest-redshift  candidates  in  the   HDF; Pascarelle, Lanzetta   \&
Fern\'andez-Soto  1998,  on the  measurements  of  the  Star Formation
History of the universe; and Fern\'andez-Soto, Lanzetta \& Yahil 1998,
on the  Luminosity  Functions as measured  from the  catalog). We will
only introduce  in this Section  some of the  basic properties  of the
galaxy distribution.

\subsection{Number counts}

  Figure \ref{nvsm} shows the  $I$-band number counts as measured from
our catalog, corrected  for  the different sizes  of the  zones Z1 and
Z2.  On   the same  diagram we  show  the  results  from some previous
studies. The number counts agree in the common magnitude range, and it
is clear from  this plot the  extension that the HDF  images represent
for this kind of  studies. It must  be noted, however, than the points
representing the work of Driver  \etal (1995) correspond only to total
counts for their   morphologically classified sample, with the  number
counts for unclassified objects reaching deeper.  In any case, because
of the small size and the particular characteristics of the field that
was chosen for the HDF, it cannot be expected to represent the general
galaxy population,  specially at the  bright  end. Driver \etal (1998)
provide a  more  in-depth study of  these  data combining the redshift
information with the morphological classification obtained by using an
Artificial Neural Network.

  We can also use the infrared images to measure number counts down to
magnitudes $J=23.5$, $H=22.5$, and  $K=22.0$. Although our  sample has
been selected in the $I$-band,  we can define  complete samples in the
infrared  given the fact that the  $I$-band image is  much deeper than
any of   the others. This  has  been checked  via an  analysis  of the
residual infrared  images, where  only one  object  brighter  than the
limits  given above  was found (in   the $K$-band image, see  Lanzetta
\etal 1998). Figure \ref{ircounts} shows  these number counts. We have
used in  this case standard magnitudes instead  of  the $AB$ system in
order to make the comparison with other published works easier.

\subsection{The redshift distribution}

  Figure \ref{nvsz} shows the redshift distribution of the galaxies in
our catalog.  We have  divided the sample  in  two different magnitude
bins (objects with  $AB(8140)<26.0$ and $26.0<AB(8140)<28.0$) in order
to show the  change of the redshift  distribution with magnitude.  The
median redshift  for the bright  magnitude  bin is $z_{\rm med}=1.00$,
and for the fainter one is $z_{\rm med}=1.56$.

  A more detailed  analysis of the redshift-magnitude distribution  is
given in Figures \ref{medz_mlim}  and \ref{cumul}. The first one shows
the change of the median redshift as a  function of apparent magnitude
limit    while the second     one shows the    cumulative (and  doubly
cumulative) distribution  of redshift for  different magnitude-limited
samples. Both figures show clearly  that the median redshift increases
as  a   function  of  the  apparent   magnitude    limit only  up   to
$AB(8140)\approx 26$,   remaining approximately  constant for  fainter
magnitudes.

\subsection{Hubble Diagram}

  Having  photometric  measurements  in seven    bands and   estimated
redshifts, we can  calculate an estimate of  the absolute magnitude of
every object in the catalog without the need to use K-corrections. The
Hubble Diagram is a simple although powerful way to  show the data and
to analyze them. Of course a more detailed analysis requires a careful
calculation  of  the  Luminosity  Function,  together   with  a proper
estimation of the  errors implied in its  determination. Nevertheless,
and for the sake of completeness,  we show in Figure \ref{hubdiag} the
rest-frame $U$-band absolute magnitude of the objects in our sample as
a function of redshift. We have  separated them according to the rough
spectral classification    given by the    best-fit  spectrum to  each
object. The choice of $U$-band magnitudes is given by the fact that we
are  calculating rest-frame magnitudes  from our best-fit spectra, and
the use of a shorter  wavelength allows for the corresponding observed
band to remain  in  the optical part of    the spectrum for  a  larger
redshift range---up to $z \approx 2$ in the case of the $U$ band, only
$z \approx  1$ for the $B$ band.   The very large  excess of Irregular
galaxies at  high redshifts might not be  completely real  because, as
was mentioned  before, at redshifts $z>3$ most  objects are  only seen
in---at  most---the  two reddest  HST bands and  are  too  faint to be
securely detected in  the IR images. In this  case the only observable
feature is  the   position of  the  Lyman Limit,  which  is  obviously
insufficient  to provide a  good estimate  of  the spectral type.  Our
algorithm tends  in this situation to choose  the Irregular SED as the
best fit simply because of the  relatively smaller flux it provides in
the   redder  filters---which  better     fits   the  data  in   every
case. Nevertheless,  once this has been remarked,  we should note that
the    morphological  classification of  galaxies  in    the HDF  with
$AB(8140)<26.0$ as  provided in  Odewahn  \etal (1996) and  studied in
conjunction with this catalog in Driver \etal  (1998) shows that there
is a real  excess of this class  of  objects at redshifts larger  than
$z=1.5$, together with a lack of  disk galaxies over the same redshift
range.

\subsection{High-redshift candidates}

  An important result from the LYF paper was the detection of galaxies
in the HDF with  estimated redshifts as high  as $z \approx 6$. One of
those  candidates (object \#213  in our  catalog, $z_{\rm phot}=5.64$)
has  already   been spectroscopically confirmed  (Weymann  \etal 1998,
$z_{\rm sp}=5.60$), and the infrared observations described in Section
2 support the high-redshift  hypothesis  for most  of them  (for  more
details, see Lanzetta, Fern\'andez-Soto \& Yahil 1998).

  First, none of the candidates shows up as a strong infrared emitter.
In case the breaks observed in  the flux of  these objects between the
$V$ and $I$  bands were due to very  strong reddening by  dust, as has
been   suggested, the spectral  energy distributions  should rise very
steeply towards the infrared---implying very red $(I-K)$ colors, which
is not observed.

  Second, the HST images alone allowed for the interpretation of those
candidates as objects in which a single extremely strong emission line
is contributing all the flux seen in  the $I$ band, with the continuum
being too weak to be detectable  in all $U$, $B$,  and $V$ bands. This
possibility  is discarded by  the fact that  several of the candidates
are detected in at least one of the infrared bands.

  Table \ref{tabhz} shows the data for the six high-redshift brightest
candidates in our catalog. All of  them have magnitudes $25 < AB(8140)
< 26.5$, which  puts them at the very  limit of feasible spectroscopic
confirmation with either Keck LRIS or STIS on  board HST. Object E has
been observed by Spinrad using  the Keck telescope  in Hawaii, and has
been {\em tentatively} assigned a redshift of $z=4.58$ (Spinrad 1997).
Figure  \ref{fighz} shows  the $UBVIJHK$ images  for  all six objects,
with the   spectral   energy  distributions and  redshift   likelihood
functions shown in Figure \ref{plothz}. From them it  can be seen that
the spectral energy distributions of objects  C, D, and  F can also be
fitted with a low-redshift solution.  Though in all cases the goodness
of fit is better for the high-redshift case,  we consider that objects
A, B and   E   are the ones    for  which the evidence favoring    the
high-redshift hypothesis is  stronger. The Weymann \etal (1998) object
is even fainter than these, and is hence not included in this list.

  Object A  is particularly  interesting   because of its   brightness
($AB(8140)=25.69$) and  its morphology.  It has  the  same colors---to
within photometric  errors---as the fainter object  to its left, which
has $AB(8140)=26.52$. In fact, if they  are taken together as a single
object in  the same way described  in Section 5, their joint magnitude
is $AB(8140)=25.30$ and the best-fit  redshift is 5.28. The reason why
both objects have not been added together in the  catalog is that they
lie in the outer zone (Z2), hence the faint one is too faint to appear
in the main  catalog. The angular  distance separating both  clumps is
0.5 arc seconds, which corresponds to 1.5  $h_{100}^{-1}$ kpc for $q_0
= 0.5$. After  this  paper had  been  submitted  Spinrad \etal  (1998)
succeeded   in measuring the   redshift  of this  faint double galaxy,
finding a  value  $z_{\rm sp}=5.34  \pm  0.01$, in excellent agreement
with the above mentioned photometric redshift.

\section{Conclusions}

  We have presented the results of a detailed 7-band photometric study
of the Hubble   Deep Field   together  with  a  catalog of   estimated
redshifts. The final catalog contains 1067 objects, is free of stellar
and point-like  objects and is complete down  to magnitude $AB(8140) =
28.0$ in   the inner 3.93   arcmin$^2$ of the  HDF  images and down to
$AB(8140)=26.0$ over the whole field.

  One  of the main  steps taken to build this  catalog is the accurate
photometry  of the IR images  provided  by Dickinson \etal (1998).  We
have  developed  an optimal  method to measure  the  IR fluxes  of all
objects detected  in the  HST   images  which doesn't rely  on   their
detection in the IR images, which allows  for the measurement of faint
fluxes in very small objects.

  We have  tested the quality of  the redshift estimates  by comparing
our results with a list of  106 spectroscopically determined redshifts
from  the literature. The  agreement   is essentially perfect  in  the
low-redshift regime ($z<1.5$),  with a $\Delta z_{\rm rms}$ dispersion
of  0.13 and  only  6\% of  the  objects  having a  difference between
photometric  and spectroscopic  redshift   larger than  0.20. At  high
redshift   ($z>2.0$) the    situation  is somehow   worse.  Our method
underestimates  the values in  the  redshift range  $z=2.0-2.5$ by  an
average $\Delta z= 0.45$. We  have a catastrophic error (i.e., $\Delta
z > 1.5$) rate of 7\% (2 out of 27)  in this range. Ignoring these two
outliers, the average dispersion is $\Delta z_{\rm  rms} = 0.45$, with
more  than 70\% of the objects  having their redshift determined to an
accuracy  of $\Delta  z  =0.50$ or  better.  The average dispersion is
significatively smaller  if   the  systematic trend observed   in  the
redshift range $z=2.0-2.5$ mentioned before is taken into account.

  We have  tested  the relative  importance  of photometric errors and
cosmic variance in producing the  observed dispersion of measurements.
Our calculations  show that cosmic  variance (i.e.,  the fact that our
set of models is only a finite sample of  all possible galaxy spectra)
is the only source of error  for objects with apparent magnitudes down
to $AB(8140)<25.0$.  Photometric errors start  to  be important in the
magnitude range $25.0 <  AB(8140) <  26.0$,  specially for  objects at
$z>2.0$.  The photometric errors alone  can account for  up to $3-6$\%
erroneous  redshift estimates in that  range.  These numbers become as
large as $10-20$\%  for  $26.0 < AB(8140) <   27.0$ and $25-35$\%  for
$27.0 < AB(8140) < 28.0$ (where the figures represent the catastrophic
error  rate in the    redshift  intervals $z=2.0-3.0$  and   $3.0-4.0$
respectively).

  The important point  that can be deduced from  this analysis is that
at the magnitudes span by the available spectroscopic redshifts cosmic
variance is the only important source  of error. Hence we can estimate
the importance of cosmic variance from  our comparison. We obtain that
it produces---over  the complete redshift range $z=0.0-4.0$---a median
error of  $\Delta z_{\rm med} \approx  0.08$,  a dispersion of $\Delta
z_{\rm   rms}  \approx 0.18$  and a  discordant   rate of 0.057. These
numbers will be  important to assess  the systematic error bars in any
future analyses to be performed using this catalog.

  We have finished our work by sketching some of the basic statistical
properties   of the ensemble  of  galaxies in   the  HDF and showing a
selection  of $z \approx 5$ candidate  objects. These galaxies as well
as the whole catalog are already the subject of subsequent studies.

\acknowledgments

  We would  like to thank  Bob Williams and  the entire STScI HDF team
for  providing the astronomical  community with such an exciting image
as the HDF is. We also thank John Webb for allowing us to use his data
on  the Lyman  $\alpha$ decrement, Mark  Dickinson for  making  the IR
images  available and Hy Spinrad,  Chuck Steidel, and  once again Mark
Dickinson  for allowing  us to     use  some of  their   spectroscopic
observations prior  to their publication. We  particularly acknowledge
the  comments from our  referee Rogier Windhorst,  that have been very
valuable  in improving the clarity of  the  paper. A.F.S. acknowledges
support from a grant from the  Australian Research Council. A.F.S. and
K.M.L.  were  supported   by  NASA  grant  NAGW--4422   and  NSF grant
AST--9624216. This  research     has made  extensive   use  of  NASA's
Astrophysics Data System Abstract Service  and the Los Alamos National
Laboratory Astrophysics e-print Archives.

\clearpage

\begin{figure}
\caption{ The zones Z1  and Z2 in  the WFPC2 images. The outer  dotted
line marks  the borders of  the 4096$^2$ mosaic  image. The continuous
line is the area  in the mosaic  covered by the  WFPC2 field, and  the
inner dashed line separates the inner Z1 and outer Z2 zones.}
\label{zones}
\end{figure}

\begin{figure}
\caption{ Comparison  of  the different  bands  and model  images. The
panels correspond to a  region of  $48  \times 48$ arcsec$^2$  in chip
4. The upper panels show the F814W and $K$-band images. The lower left
panel is the $K$-band model image as described on  the text. The lower
right corner shows the image resulting from subtracting the model from
the original $K$-band image.}
\label{image}
\end{figure}

\begin{figure}
\caption{ Distribution   of values  of  the parameter   $q$ in  all IR
bands. From  left to right  the  distributions correspond to  the $J$,
$H$,  and  $K$ band  images.   The   superposed continuous lines   are
Gaussians with zero mean value and unit sigma, fitted to have the same
number  of  objects as   each of the   distributions  in the  negative
part. Observe the broadening of the positive tails of the distribution
due to low-flux real objects.}
\label{errors}
\end{figure}

\begin{figure}
\caption{   Analysis of the  residuals   after model subtraction.  The
dotted lines show the distribution  of pixel intensities in the  areas
of  every  image  that  do not   correspond to  any object (1-$\sigma$
criterium).  The continuous lines show  the same  distribution for the
pixels  that lie within objects. The  distributions are very different
in the original images (upper panels),  but do agree sensibly once the
model  image is subtracted  from the original  (lower panels), showing
the   goodness of the fitting  procedure.   Note that the dotted lines
(background) are the same in the upper and lower panels.}
\label{residuals}
\end{figure}

\begin{figure}
\caption{ Spectral  Energy    Distributions used  for  the    redshift
estimation procedure. The  models    starting  from the   lowest   one
correspond to E/S0, Sbc, Scd, and Irr galaxies.}
\label{sed}
\end{figure}

\begin{figure}
\caption{ Comparison of photometric and spectroscopic redshifts.}
\label{spvsph}
\end{figure}

\begin{figure}
\caption{ Comparison of  photometric and  spectroscopic redshifts. The
objects have been binned in groups of nine or  ten per bin. The points
correspond  to the median redshift  and value of $\Delta  z  / (1+z) =
(z_{\rm sp}-z_{\rm phot}) / (1+z)$ per  bin, with the horizontal error
bars showing the width of the bin and the  vertical error bars showing
the 25\% and 75\% percentiles in each bin.}
\label{delta}
\end{figure}

\begin{figure}
\caption{  Effect  of   the   photometric  errors in     the  redshift
determination procedure. Each box shows  the result of the simulations
for a given redshift and  magnitude interval as labeled, together with
the median redshift difference  obtained.  The graphic shows how   the
influence of the   photometry   becomes  more  important at    fainter
magnitudes,  with both the median  difference and  the secondary peaks
centered at the wrong redshift each becoming larger.}
\label{boots}
\end{figure}

\begin{figure}
\caption{  Galaxy number counts.  The filled symbols correspond to our
catalog of  the  HDF images.  The empty  circles  show the  results of
Casertano \etal  (1995), the empty squares  those of Driver, Windhorst
\& Griffiths  (1995), and the  empty triangles  those of  Driver \etal
(1995). The  latter  are only plotted  to  the magnitude  at which the
authors  could morphologically classify the  objects from their images
($I < 24.5$).}
\label{nvsm}
\end{figure}

\begin{figure}
\caption{  Galaxy number  counts in the  $J$,  $H$, and $K$-bands. For
clarity the counts in $J$ and  $H$ have been displaced  up two and one
orders of magnitude respectively. The vertical lines mark the position
of the 5$\sigma$  completeness limits. No completeness  correction has
been applied below these levels.}
\label{ircounts}
\end{figure}

\begin{figure}
\caption{ Galaxy  redshift   distribution from  our  catalog. We  have
broken the catalog   in two magnitude-divided  subsamples in  order to
show the change in the redshift distribution with magnitude.}
\label{nvsz}
\end{figure}

\begin{figure}
\caption{ Median redshift as a function  of magnitude limit. For every
magnitude bin, the filled triangle and the dashed line mark the median
redshift  and  68\% interval around  it for  all galaxies  within that
redshift bin. The filled circle with the continuous line mark the same
for the   sample of galaxies  with  a magnitude  flux in  that  bin or
brighter. The positions of  the different points  and lines  have been
slightly offseted for clarity.}
\label{medz_mlim}
\end{figure}

\begin{figure}
\caption{ Upper panel: Cumulative redshift distribution of galaxies in
different  magnitude  bins.  Each line   shows  the percentage of  the
galaxies within the magnitude range $(m-1,m)$---starting at $m=22$ and
moving towards the right to $m=28$--- that have redshift below a given
value. Lower panel:  Same but applied to  the sample of  galaxies with
magnitude  brighter than   $m$.  In  both  cases the  lines  alternate
continuous and dotted style  for  clarity and  the ticks close  to the
X-axis mark the  median redshift  for  each of  the  magnitude-limited
samples.}
\label{cumul}
\end{figure}

\begin{figure}
\caption{ The Hubble diagram for the galaxies  in our catalog. Each of
the panels shows the diagram for each of  the galaxy types assigned by
our method. The dotted lines mark the detection  limits set on the HDF
images ($18.0 < AB(8140) < 28.0$) taking into account the SED of every
galactic type.}
\label{hubdiag}
\end{figure}

\begin{figure}
\caption{ $UBVIJHK$ band   images of our six brightest   high-redshift
candidates. Each box is $5''\times5''$ in size.}
\label{fighz}
\end{figure}

\begin{figure}
\caption{ Spectral    energy  distributions  and  redshift  likelihood
functions  for our six  brightest high-redshift candidates.  The solid
dots  mark the photometric   data,  while the   open circles show  the
expected response  from the best-fit  model galaxy templates. Vertical
error bars correspond to  photometric errors and horizontal error bars
indicate the FWHM of the filters.}
\label{plothz}
\end{figure}

\clearpage

\begin{table}
\caption{\label{psf} Parameters for  the quadratic  fit of the   point
spread function FWHM in  every band (FWHM$_{pixels}=a+br^2$, where $r$
= distance to the center of the image in pixels).}
\begin{tabular}{ccc} \tableline \tableline
Band & $a$ & $b$ \\ \tableline
 $J$ & 6.4009 & $3.454 \times 10^{-6}$ \\ 
 $H$ & 6.2030 & $4.498 \times 10^{-6}$ \\ 
 $K$ & 6.1367 & $3.722 \times 10^{-6}$ \\ \tableline
\end{tabular}
\end{table}

\clearpage

\begin{table}
\caption{\label{galparam} Ages,  types  of Initial  Mass Function  and
Star Formation  Rates used by the  Bruzual  \& Charlot  (1993) code to
generate the  IR part of the model  galaxy spectra. In the last column
$\tau_n$ denotes  an exponentially  decaying  SFR  with characteristic
time of $n$ Gyr.}
\begin{tabular}{cccc} \tableline \tableline
    Type &  Age (Gyr) & IMF &  SFR \\ \tableline
 E/S0 &  12.7  & Scalo &  $\tau_1$ \\
 Sab/Sbc & 12.7 & Scalo & $\tau_8$ \\
  Scd & 12.7 & Salpeter & Constant \\ 
   Irr & 0.1 & Salpeter & Constant \\ \tableline
\end{tabular}
\end{table}

\clearpage

\begin{table}
\caption{\label{stars} Stars in the Hubble Deep Field. The coordinates
refer to $(x,y)$ positions in the Mosaic images.}
\begin{tabular}{c c c c} \tableline \tableline
\multicolumn{2}{c}{Coordinates} & AB(8140) & References\tablenotemark{a} 
\\ \tableline
 1877 &  349 & 23.52 & 1,2 \\
 3380 &  404 & 22.68 & 1,2 \\
 1306 &  420 & 24.64 & 1,2,4 \\
  869 &  988 & 19.22 & 1,4 \\
  247 & 1117 & 22.50 & 1,2,3,4 \\
 2904 & 1130 & 21.08 & 1,4 \\ 
 1194 & 1254 & 22.71 & 1,2,3,4 \\
 2449 & 1574 & 22.40 & 1,2,3,4 \\
 2070 & 1756 & 20.89 & 1 \\
 1026 & 1802 & 21.23 & 1,2,3,4 \\
 3425 & 1920 & 24.14 & 1,2,3,4 \\
 2441 & 1936 & 24.34 & 1,2,3,4 \\
  654 & 2180 & 19.26 & 1,4 \\
 1645 & 2496 & 27.61 & 3 \\
  535 & 2775 & 26.76 & 1,2,3 \\
  956 & 2845 & 25.42 & 1,2,3 \\
 1095 & 3203 & 26.38 & 1,2 \\
  315 & 3788 & 20.84 & 1,3 \\
 1389 & 3819 & 25.50 & 3 \\
 1361 & 3974 & 19.90 & -- \\ \tableline
\end{tabular}
\tablenotetext{a}{The  references  are:  1)  Flynn,  Gould \&  Bahcall
(1996),  2) Elson, Santiago   \&  Gilmore (1996),  3)  M\'endez  \etal
(1996), 4) Cowie (1997). The last entry is an obvious star not present
in any  of the references,  perhaps because it lies   too close to the
upper border of the image.}
\end{table}

\clearpage

\begin{table}
\setcounter{table}{5}
\caption{\label{tabhz} Data    corresponding    to   the     brightest
high-redshift candidates in our catalog.}
\begin{tabular}{crrrrrcc} \tableline \tableline
Name & No\tablenotemark{a} & \multicolumn{4}{c}{Coordinates
\tablenotemark{b}} & $z_{\rm phot}$  & $AB(8140)$ \\ \tableline
 A &    3 &  542.8 &  229.2 & 12:36:59.797 & 62:12:18.67 & 5.72\tablenotemark{c} & 25.69 \\
 B &  311 & 2342.2 &  968.4 & 12:36:48.714 & 62:12:16.72 & 5.64 & 26.32 \\
 C &  328 & 2393.8 & 1006.4 & 12:36:48.357 & 62:12:17.28 & 4.76 & 26.19 \\
 D &  868 &  407.6 & 2752.9 & 12:36:54.714 & 62:13:52.85 & 4.72 & 25.97 \\
 E &  173 & 3259.2 &  652.1 & 12:36:44.658 & 62:11:50.45 & 4.58\tablenotemark{d} & 25.04 \\
 F &  479 &  887.0 & 1392.1 & 12:36:55.334 & 62:12:55.54 & 4.56 & 25.24 \\ \tableline
\end{tabular}
\tablenotetext{a}{Number in our catalog} 
\tablenotetext{b}{$(X,Y)$ in the mosaic images and (RA,DEC)}
\tablenotetext{c}{Or $z_{\rm phot}=5.28$, see text for explanation. $z_{\rm   
spec}=5.35$}
\tablenotetext{d}{Tentative spectroscopic redshift (see text), $z_{\rm
phot}=4.52$}
\end{table}

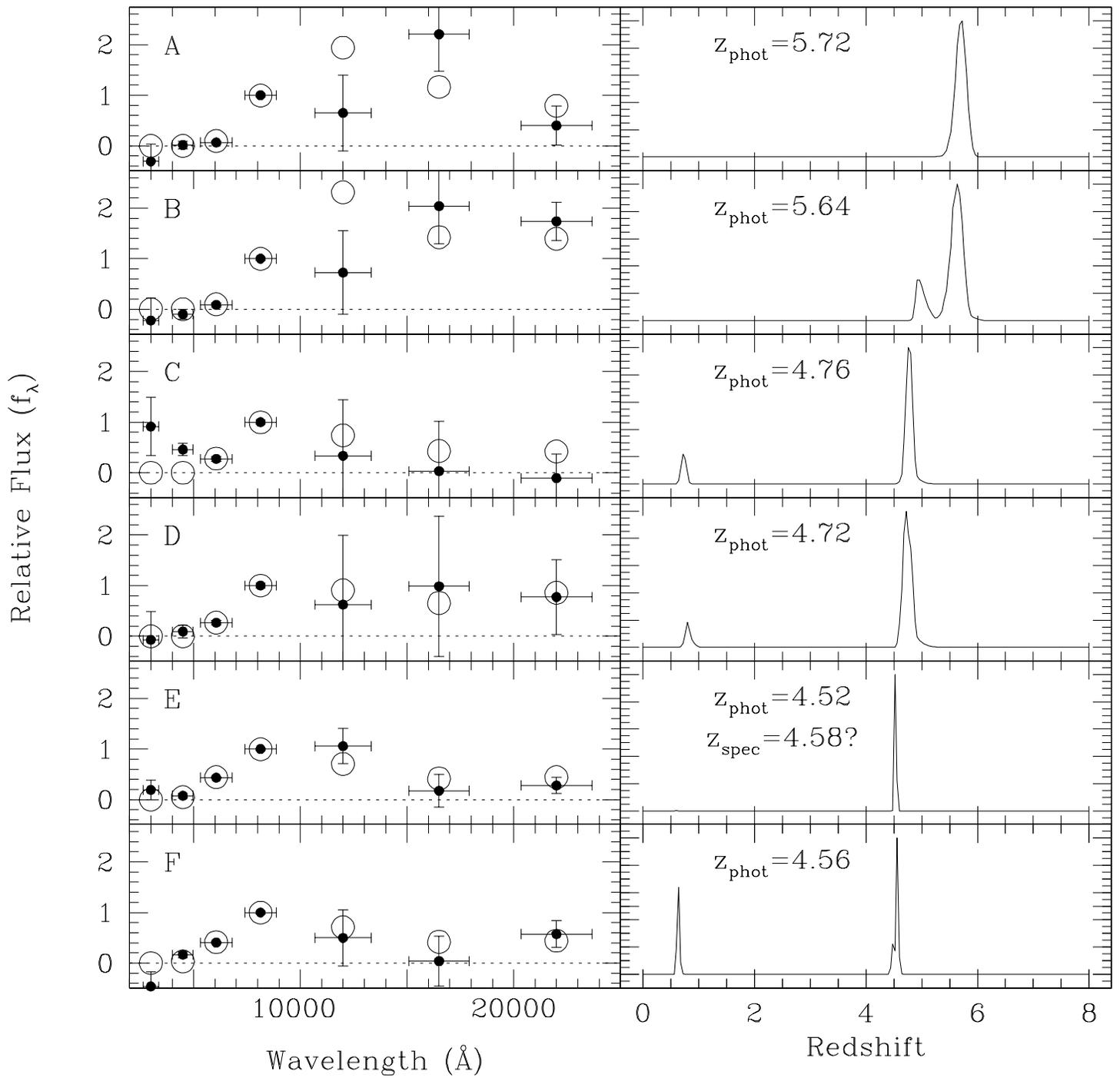
\end{document}